\documentclass[9pt,twocolumn,twoside]{opticajnl}
\journal{opticajournal} 

\setboolean{shortarticle}{false}

\usepackage{siunitx}%
\usepackage{braket}
\usepackage{lineno}

\title{Speckle Angle: An Intrinsic Geometric Metric for High-Precision Speckle Metrology}

\author[1,2,*]{Alexander C Trowbridge}
\author[1,2,3]{Gabriel Britto Monteiro}
\author[1,2,3,4]{Kishan Dholakia}

\affil[1]{Centre of Light for Life, School of Biological Sciences, Adelaide University, Adelaide, 5005, SA, Australia}
\affil[2]{Institute for Photonics, Advanced Sensing and Quantum Technologies, Adelaide University, Adelaide, 5005, SA, Australia}
\affil[3]{ARC Centre of Excellence in Optical Microcombs for Breakthrough Science (COMBS), Adelaide University, Adelaide, 5005, SA, Australia}
\affil[4]{SUPA, School of Physics \& Astronomy, University of St Andrews, North Haugh, St Andrews, KY16 9SS, UK}

\affil[*]{alexander.trowbridge@adelaide.edu.au}

\begin{abstract}
Speckle-based metrology exploits the extreme sensitivity of complex interference patterns to perturbations in wavelength, refractive index, displacement and other physical parameters. Existing approaches typically quantify speckle evolution using similarity-based correlation metrics, whose local sensitivity deteriorates near the operating point because their gradient vanishes as the similarity approaches unity. Here we show that normalized speckle patterns represent points on the surface of a high-dimensional unit sphere and introduce the speckle angle: the intrinsic geometric coordinate between speckle patterns. The speckle angle preserves the physical calibration of established similarity approaches while providing a linear response and maximal sensitivity in the small-perturbation regime. This geometric perspective enables discrimination of perturbation dynamics from measurement noise, enabling direct estimation of uncorrelated and fixed-pattern noise contributions and yielding a fundamental acquisition-limited resolution bound. Using speckle generated by an integrating sphere, we experimentally validate the framework and demonstrate quantitative noise analysis using angular coordinates. Beyond improving sensitivity characterization, the speckle angle establishes an original common metric for comparing disparate speckle metrology architectures and provides a foundation for multidimensional sensing based on trajectories through speckle-vector space.

\end{abstract}

\setboolean{displaycopyright}{false} 

\begin{document}

\maketitle

\section{Introduction}
Speckle metrology has emerged as a powerful sensing paradigm capable of resolving extraordinarily small variations in wavelength, refractive index, displacement, and other physical parameters \cite{bruce_overcoming_2019, facchin_measurement_2022, facchin_determining_2024, perrella2026speckle, britto2026ultra}. In a typical implementation, the response of a speckle-generating system is quantified by comparing speckle patterns acquired at different values of a measurand \cite{facchin_determining_2024}. While the spectral correlation function remains the most widely used metric for this purpose \cite{redding2012using, redding2013all, redding2014high, cao2022harnessing, bruce_overcoming_2019, zhang2025scalable, wan2021review, french2018snapshot, ruffing1986application}, similarity-based approaches are attractive because of their ability to support high-precision measurements \cite{chakrabarti2015speckle, trivedi20193d, facchin_wavelength_2021, tran2020utilizing, facchin_measurement_2022, facchin_measuring_2023, trivedi2022opto}.

A key advantage of the similarity metric is its direct connection to physical changes in path-dependent measurands, enabling an intuitive and quantitative interpretation of speckle evolution \cite{facchin_determining_2024, facchin_wavelength_2021}. However, similarity is not the natural coordinate of speckle evolution, it is a projection. The similarity response is also typically nonlinear and approaches unity at the operating point. Consequently, its local gradient becomes vanishingly small in this regime, reducing the sensitivity of the metric to incremental changes in the measurand. These limitations present significant challenges for widespread adoption and use of the similarity in precision speckle metrology experiments. This is exemplified by our recent study, which used speckle from an integrating sphere to measure the refractive index of a gas \cite{facchin_measurement_2022}. Though this limited sensitivity was circumvented in this study through careful experimental design, it required simultaneous control of multiple independent variables, complicating the experiment. Additionally, rigorous characterization of the noise floor and ultimate measurement resolution remains difficult despite the physical interpretability of the similarity formalism. These challenges are further compounded by the high-dimensional representation of speckle data.

Although the underlying speckle field contains a finite number of independent optical modes, modern imaging systems encode these measurements on a camera array comprising millions of pixels. Increasing detector resolution can provide access to additional spatial information, but simultaneously increases the dimensionality of the recorded dataset. Extracting metrologically relevant information from such measurements therefore remains a central challenge in speckle-based sensing. To address this issue, dimensionality-reduction techniques, most notably principal component analysis (PCA), have been widely adopted to identify and track the dominant modes of speckle variation \cite{metzgerHarnessingSpeckleSubfemtometre2017, perrella2026speckle, bhagat2026non, bajraktarevic2026enhancing}. Although highly effective at compressing large datasets while preserving measurement-relevant information, PCA does not, in general, provide an explicit physical interpretation in terms of the measurand. Consequently, the extracted components typically require empirical calibration for each measurement scenario and, in some cases, may preferentially capture other sources of variance, including environmental fluctuations or measurement noise. This disconnect between statistically significant variations and physically meaningful changes presents a fundamental limitation for quantitative speckle metrology. For metrological applications, a direct correspondence between the analysis metric and the measurand is highly desirable because it simplifies both noise-floor estimation and physical interpretation. Without such a physics-grounded framework, assessments of sensitivity and resolution remain dependent on dataset-specific calibration and careful post hoc validation. These considerations motivate the development of physically interpretable geometric observables that retain sensitivity near the operating point while providing a rigorous framework for noise-floor quantification and uncertainty analysis.

Here, we show that the speckle similarity formalism admits a simple yet powerful geometric interpretation that exposes it as a projection of an underlying, more fundamental degree of freedom. Normalized speckle intensities may be represented as points on the surface of a unit $N$-sphere, such that their evolution is naturally quantified by the angular separation between points on this manifold. We term this quantity the speckle angle. In contrast to conventional similarity metrics, the speckle angle exhibits a linear response and maximal sensitivity in the small-perturbation regime while retaining direct compatibility with established calibration procedures. Beyond providing a more intuitive description of speckle dynamics, this geometric framework enables the tracking of independent trajectories on the speckle manifold, opening a route toward the simultaneous extraction of multiple measurands from a single evolving speckle field. We perform a case study of this metric using speckle from an integrating sphere, demonstrating its utility in the small perturbation regime. In spite of this, the geometric framework that presented here is independent of the scattering system and applies equally to other scattering media, including multimode optical fibers, disordered photonic structures, and rough surfaces.

\section{Similarity}
Similarity provides a convenient means of relating speckle patterns to changes in an underlying physical parameter. Although originally formulated in the context of wavelength perturbations, the concept has since been generalized to a wide range of path-dependent measurands \cite{chakrabarti2015speckle, trivedi20193d, facchin_wavelength_2021, tran2020utilizing, facchin_measurement_2022, facchin_measuring_2023, trivedi2022opto}. The specific choice of measurand is not essential to the formalism. We therefore consider a generalized nondimensional perturbation, \begin{equation} \Delta=\frac{\Delta\alpha}{\Delta\alpha_0}, \end{equation} where $\Delta\alpha_0$ denotes a characteristic scale of the measurand. Let $I$ denote the intensity array recorded by a camera comprising $N$ pixels. We define the speckle vector as a normalized intensity,

\begin{equation} \label{eq:vec_transform}
|i\rangle \equiv \lim_{N\rightarrow\infty} \frac{1}{\sqrt{N}} \frac{I-\langle I\rangle}{\sigma_I}, 
\end{equation} 

where $\langle I\rangle$ and $\sigma_I$ are the spatial mean and standard deviation of the recorded intensities, respectively. By construction, $\ket{i}$ is a real unit vector. The bra-ket notation is introduced to emphasize that a speckle pattern should be regarded as an element of an underlying Hilbert space rather than as a finite collection of pixel values. Increasing the detector resolution refines the sampling of the same physical object, and thus vector operations between sufficiently sampled speckle patterns converge toward detector-independent quantities. Therefore, any experimentally measured and normalized intensity array represents a finite-dimensional approximation to the abstract vector $|i\rangle$. Within this framework, the similarity between two speckle patterns, represented by the vectors $|i\rangle$ and $|j\rangle$, is simply the inner product 
\begin{equation} S=\braket{i|j}= f(\Delta). \end{equation}
For real normalized vectors, this is equal to the cosine of the angle between them and is given by some function, $f(\Delta)$, that is specific to the scattering medium. Similarity is a useful metric when the perturbation is comparable to the characteristic measurand scale (i.e. $\Delta\simeq1$).

\begin{equation}
    S \equiv \braket{i|j} = \cos\theta.
    \label{eq:similarity_def}
\end{equation}

However, in practice we rarely wish to consider the similarity between patterns, rather the difference. This is a similar concept, but reframing it in this way has some important implications. The similarity has a zero gradient around $\Delta=0$. Consider the \emph{distance} between the two speckle vectors. That is, the magnitude of the difference between the two. In a geometrical interpretation, this is related to the sine of the aforementioned angle, which is maximally sensitive around $\Delta=0$ in the ideal case. With some manipulation, one can show $D^2 = 2(1-S)$.

\begin{equation}
    D = ||\,\ket{i} - \ket{j}\,|| = 2\sin\frac{\theta}{2}.
    \label{eq:distance_def}
\end{equation}

\section{The angle}
\begin{figure}[htbp]
    \centering
    \includegraphics[width=0.7\linewidth]{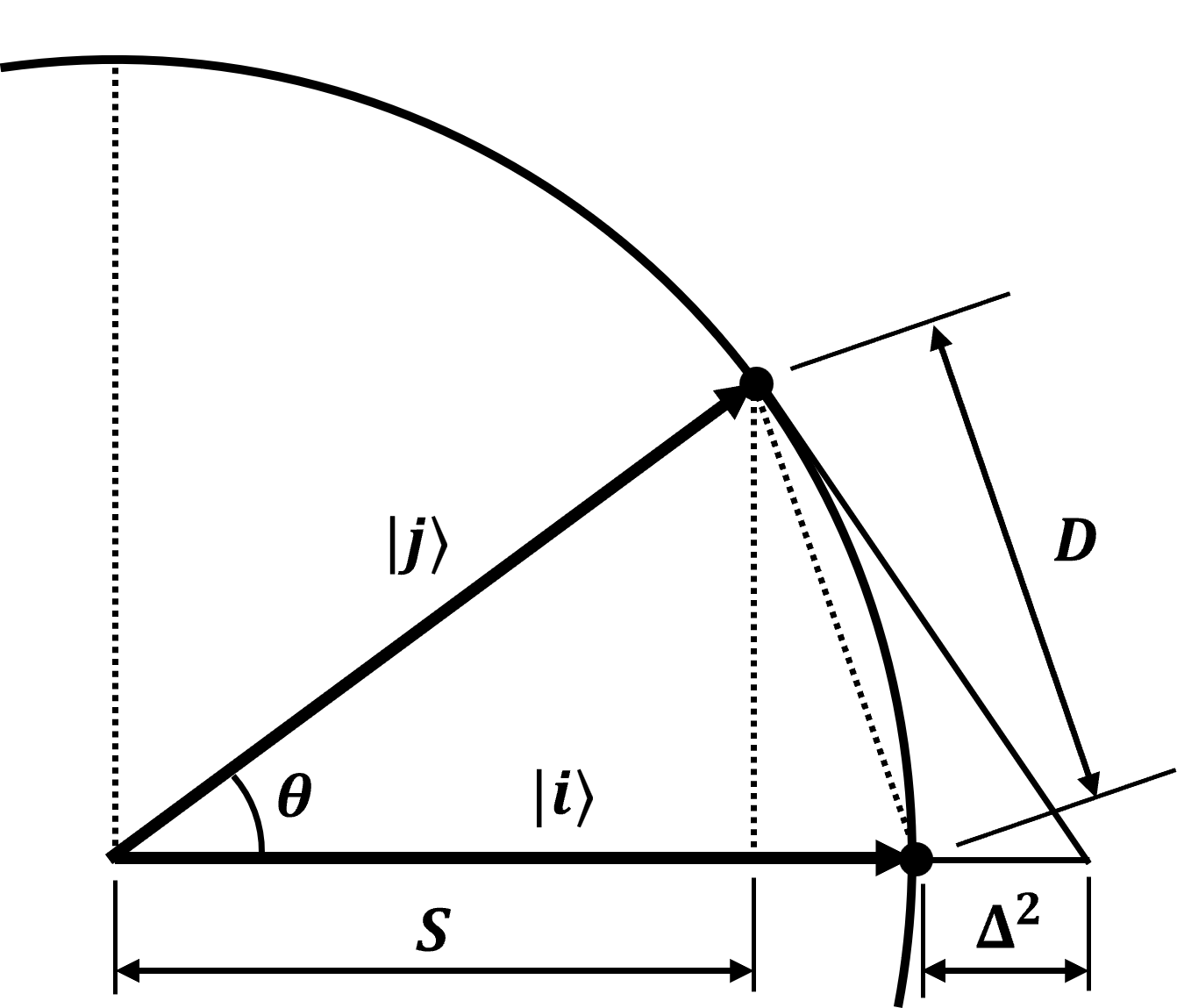}
    \caption{Normalized intensity patterns may be represented as points on a unit sphere in a high-dimensional vector space by means of corresponding speckle vectors. The angle between vectors $|i\rangle$ and $|j\rangle$ provides a natural measure of pattern evolution between measured intensity patterns $I_i$ and $I_j$. Similarity $S$ corresponds to the projection of one vector onto another, whereas speckle angle directly quantifies the natural degree of freedom distinguishing the patterns. This diagram represents a 2-D cross section of the unit $N$-sphere spanned by the two vectors. Note, the length $\sec\left(\theta\right) - 1$ is labelled $\Delta^2$ as it may be directly related to the perturbation, $\Delta$, when an integrating sphere is used to generate laser speckle. This is introduced explicitly in \autoref{eq:sphere_angle}.}
    \label{fig:angle}
\end{figure}

Both of these parameters are related by $\theta$; this is what we refer to as speckle angle. In the small perturbation regime it approximately equal to the distance between speckle patterns. The geometric interpretation of the angle for two speckle vectors $|i\rangle$ and $|j\rangle$ is depicted in \autoref{fig:angle}. As stated above, the similarity can be interpreted as the cosine of the angle. From there, all other related physical quantities inherit geometric aspects.

The angle therefore constitutes an intrinsic geometric coordinate describing the angular separation of a speckle pattern from a reference pattern. Unlike similarity, which measures a projection onto the reference vector, the speckle angle directly quantifies movement along the surface of the unit $N$-sphere manifold. Although we argue here that the angle is the natural coordinate for the comparison of speckle patterns, it remains best calculated from captured intensity patterns via similarity or distance, depending on the perturbation scale. The similarity provides a means of calibrating changes in speckle patterns over large perturbations, while the distance allows the calibration to be applied to small perturbations, or to noise analysis.

\begin{equation}
    \theta = \arccos(\braket{i|j} ) = 2\arcsin\left( \frac{||\,\ket{i} - \ket{j}\,||}{2}\right).
    \label{eq:angle_def}
\end{equation}

While the speckle angle introduced here measures the separation between speckle patterns, the associated displacement vector in the tangent space contains directional information. Different physical perturbations are expected to generate distinct tangent directions. This suggests a route towards simultaneous discrimination of multiple measurands. Exploration of the metrological implications of this higher dimensional geometric structure is the topic of future work.

\section{Noise analysis}
Ideally, a speckle experiment maximises signal to noise. However, in real experiments there will be intensity contributions from signal, $s$, uncorrelated noise, $u$ and correlated noise, $c$. Uncorrelated noise differs unpredictably between patterns. Examples include laser shot noise and dark noise from the camera sensor. On the other hand, correlated noise is constant between patterns, examples non-ideal speckle contrast, and fixed pattern noise from the camera sensor~\cite{peterkovicOptimizingImageCapture2025}. Here we express overall captured intensity pattern as the sum $I = I^{(s)} + I^{(u)} + I^{(c)}$. The vector corresponding to the overall captured intensity can be expressed as $\ket{i} = \beta_s\ket{s} + \beta_u\ket{u} + \beta_c\ket{c}$, where each $\beta$ corresponds to the relative strength of the intensity contributions such that their quadrature sum is unity.

By construction, the normalised unit vector counterparts of these vector components obey the following relations: $\braket{u_i | u_j}=\delta_{ij}$, $\braket{c_i | c_j}=1$, $\braket{s_i | s_j}=\cos{\theta}=f(\Delta)$. For speckle patterns with a larger number of resolved grains, all intensity components are approximated as orthogonal. Even with this simplified intensity model, both uncorrelated and correlated noise contributions become evident.

\begin{align}
    \theta &= \arccos(\braket{i|j}) \\ &= \arccos\left( \beta_c^2 + \beta_s^2f(\Delta) \right) \;\mathrm{for}\,\; i\neq j.
\end{align} \label{eq:theta_noise}

The limits of small and large perturbations show the existence of a departure angle, $\theta_0 = \arccos(1-\beta_u^2)$ and an asymptote angle, $\theta_\infty = \arccos\beta_c^2$, respectively. This gives the relative intensity fractions for uncorrelated and correlated noise to be $\beta_u=D_0/\sqrt{2}$ and $\beta_c=\sqrt{S_\infty}$ respectively, where $D_0$ and $S_\infty$ are the initial distance and similarity asymptote, respectively. Both $D_0$ and $S_\infty$ are easily measured experimentally, providing a means of analysing noise contributions of experimental data without further analysis. Although it is technically possible to rely on the similarity alone for characterising noise noise, it complicates the analysis. For example, the uncorrelated noise component appears only in extremely low SNR scenarios. Additionally, in high signal to noise measurements, such as in \cite{facchin_wavelength_2021}, the similarity appears to return to unity masking the contribution of uncorrelated noise entirely. This arises from the insensitivity of $S$ to $\Delta$ for small perturbations. Bajraktarevic \textit{et al.} discuss these limitations in more detail \cite{bajraktarevic2026enhancing}.

Shot noise is an important uncorrelated noise source. It is given by $\sigma_{SN}=\sqrt{n}$ for $n$ photons per pixel. For the sake of presenting a lower bound, we assume there is no correlated noise, $\beta_c=0$. This yields a signal to noise ratio of $\beta_s/\beta_u=\sqrt{\langle n\rangle}$ across the intensity pattern. After some minor manipulation, one may show that the shot noise angle bound is given by,

\begin{equation}
    \theta_{0,\,SN}=\arccos\left(\frac{\langle n\rangle}{\langle n\rangle+1}\right).
\end{equation}

For the typical regime of significant photon counts ($n>>1$) but small angles, the angle bound approximates to $\theta_{0,SN}\simeq \sqrt{2/\langle n \rangle}$.

The gradient of the angle is of particular interest when considering the sensitivity of the parameter. In the absence of noise contributions, the angle is most sensitive at zero perturbation. However, if any uncorrelated noise is present in the images, there will exist a region for which changes in the measurand are overwhelmed by this noise. The edge of this perturbation region, denoted $\Delta\alpha^*$, corresponds to the value of $\Delta\alpha$ which maximises the derivative of $\theta$ with respect to $\Delta\alpha$.

\begin{equation} \label{eq:max_sens}
    \Delta\alpha^*=\arg\max_{\Delta\alpha}\frac{\partial\theta}{\partial\left(\Delta\alpha\right)}
\end{equation}

With some manipulation it can be shown this expression is satisfied by the following differential equation in the measurand function, $f(\Delta)$.
\begin{equation}
    0 = f^{\prime\prime} + \beta_s^2 \frac{\beta_c^2+\beta_s^2 f}{1 + (\beta_c^2+\beta_s^2 f)^2} f^{\prime2}
\end{equation}
where $f^\prime$ and $f^{\prime\prime}$ represent the first and second derivatives of $f(\Delta)$ with respect to $\Delta$.

\section{Experimental validation}\label{subsec:exp}



We conducted a case study using laser speckle from an integrating sphere. An integrating sphere was chosen because there exists an analytical form for the similarity dependence on a change in wavelength in the absence of noise, $\Delta\lambda$ \cite{facchin_wavelength_2021}: 
\begin{equation}
     f(\Delta) = S_{\beta_s=1} = \frac{1}{1 + \Delta_\lambda^2};\quad \Delta_\lambda = \frac{\Delta\lambda}{\Delta\lambda_0},
\end{equation}
where $\Delta\lambda_0 = 3\lambda^2|\ln\rho|/(8\pi R)$ and: $\lambda$ is the wavelength from which the change is measured (i.e. the laser wavelength), $R$ is the radius of the sphere, and $\rho$ is the reflectance of the internal surface of the integrating sphere. Using this relationship between the perturbation and the angle can be shown to follow the expression,
\begin{equation}
    \Delta_\lambda^2=\frac{\beta_s^2}{\cos\theta-\beta_c^2}-1 \;\mathrm{or}\; \Delta_\lambda^2 = \sec\theta - 1 \;\text{without noise}.
    \label{eq:sphere_angle}
\end{equation}

\begin{figure}[htbp]
    \centering
    \includegraphics[width=\linewidth]{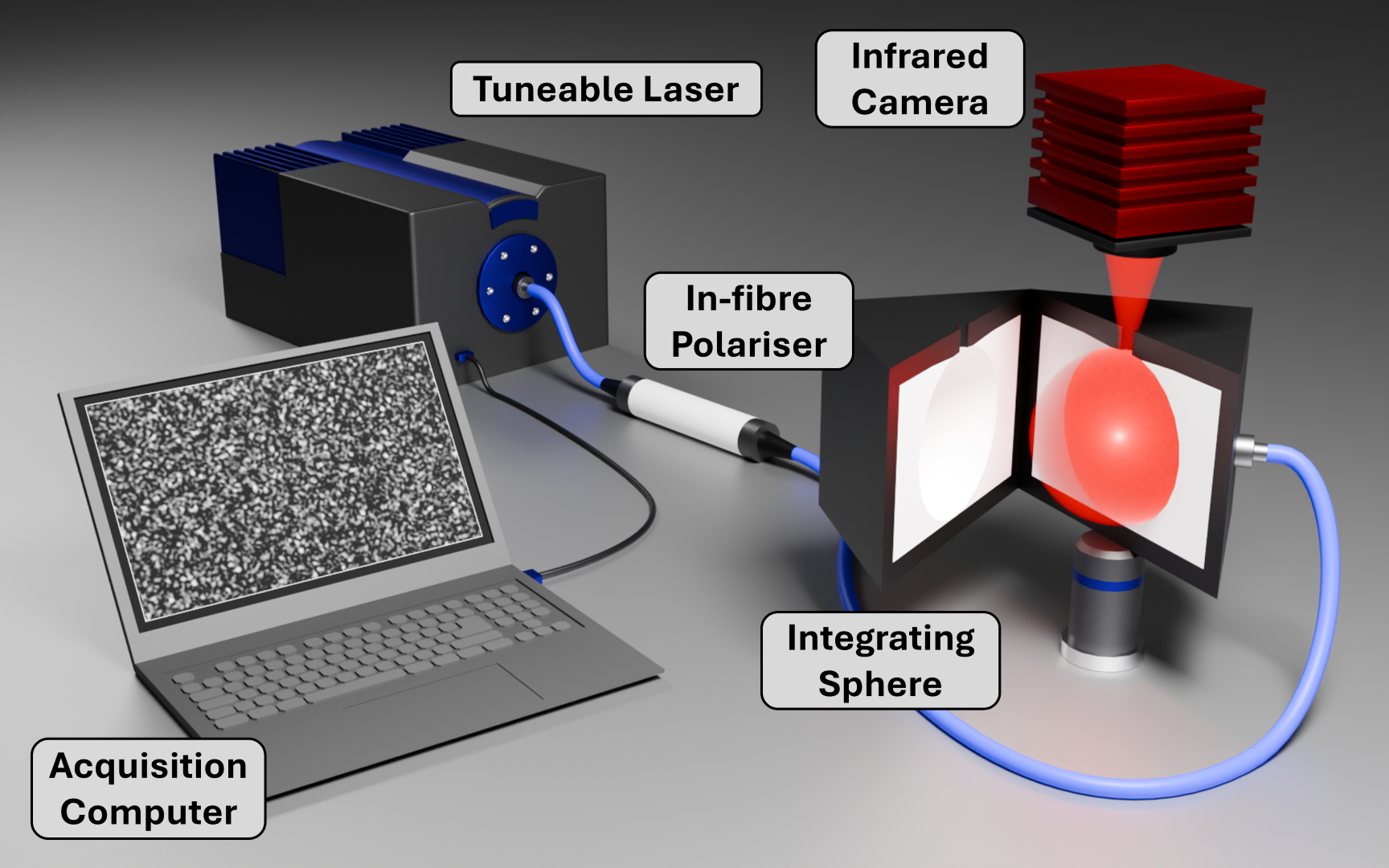}
    \caption{Experimental setup. Light from a fibre-coupled tuneable laser passes through an in-fibre linear polariser before being coupled into an integrating sphere. Light exiting the integrating sphere from its output port forms a speckle pattern and is captured on a short-wave infrared (SWIR) camera. A local personal computer controls the laser and stores digital speckle patterns from the SWIR camera.}
    \label{fig:experiment}
\end{figure}

To conduct the case study, a tuneable laser (Agilent, 81640A) was fibre coupled to a 10\,mm diameter polymer integrating sphere (Artifex Engineering GmbH, 20.220.00076) using polarization maintaining fibre. Before launching into the sphere, the laser light passed through an in-fibre linear polarizer (AFW Technologies, ILP-15-L-1-7-2-SA). Light within the integrating sphere was multiply scattered before exiting via the output port, 3.5\,mm in diameter. A short-wave infrared camera (First Light Imaging, C-RED2) captured speckle patterns emanating from the output of the integrating sphere. The camera was placed approximately 10\,cm away from the output of the sphere, yielding speckle grains that were on average $3\times 3$ pixels in the horizontal and vertical directions, respectively. Both the integrating sphere and camera were housed inside a homemade passive acoustic isolation chamber. Before data collection, light was injected into the sphere and both the sphere and camera were allowed to equilibrate within the isolation chamber. A schematic for this experiment is shown in \autoref{fig:experiment}.

The wavelength of the tuneable laser was swept linearly from 1550\,nm to 1551\,nm at $0.5\,$nm/s using the inbuilt function on the laser controller (Agilent, 8164A). During the wavelength sweep, the short-wave infrared (SWIR) camera captured speckle patterns using an integration time of 0.28\,ms. The camera region of interest was set to $640\times 64$ 14 bit pixels, corresponding to approximately 4550 spatial channels and enabling a frame rate of 3486\,Hz. Speckle patterns were extracted from the camera after the wavelength sweep and were stored on a computer for analysis. After the wavelength sweep, a mechanical shutter blocked light leaving the sphere. Subsequently, 5000 dark frames were captured. These were used to demonstrate the influence of correlated noise on the angle.

\begin{figure}[htbp]
    \centering
    \includegraphics[width=\linewidth]{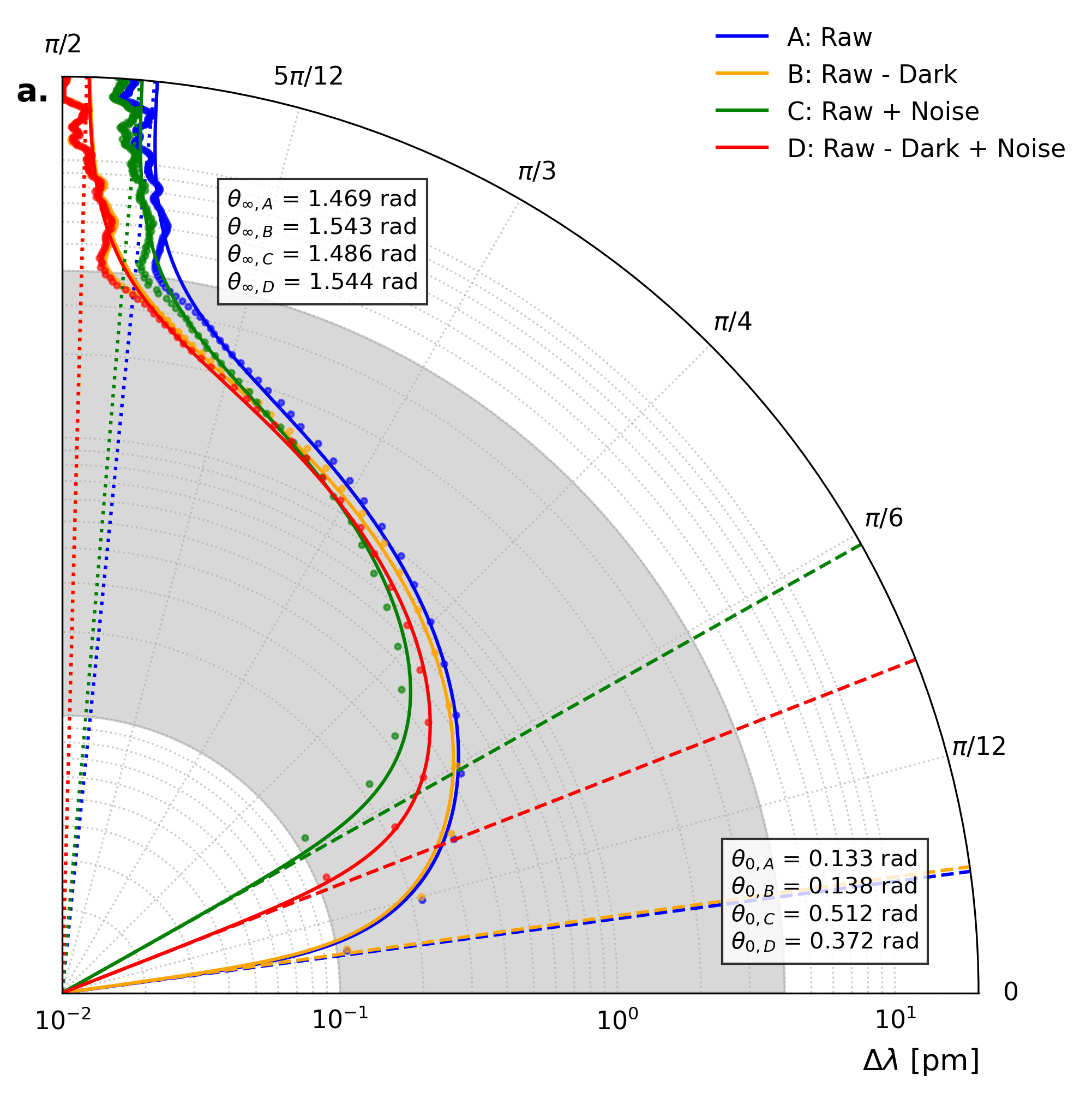}
    \includegraphics[width=\linewidth]{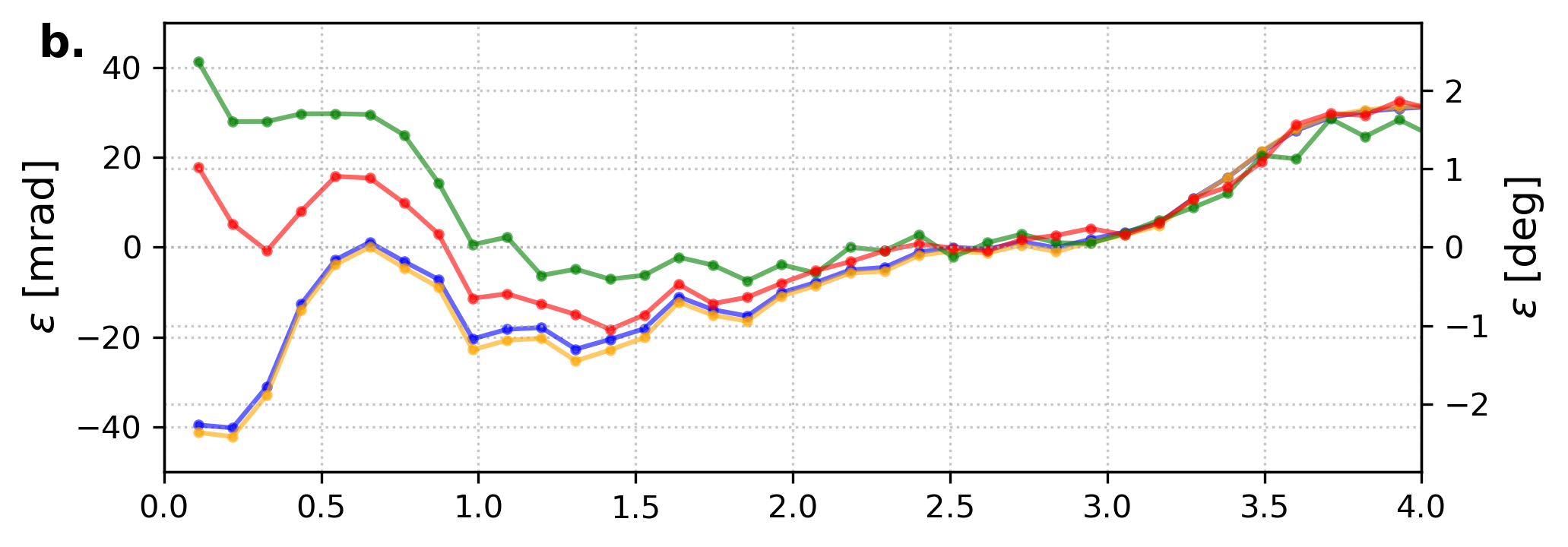}
    \caption{Speckle patterns were captured as the wavelength of a tuneable laser was swept from 1550\,nm to 1551\,nm. Coloured markers represent experimental data while solid lines represent corresponding lines of best fit to \autoref{eq:angle_def}. The plot shows four traces: the raw data, dark background subtracted, added Gaussian noise, and a combination. This allows one to see clearly the influence of both uncorrelated and correlated noise. \textbf{a.} Experimental trace of speckle angle for the setup of \autoref{fig:experiment}. The wavelength change, $\Delta\lambda$, was cropped to a reduced range of 20\,pm and is plotted on a logarithmic axis. The departure and asymptote angles for each dataset are shown using dashed and dotted lines, respectively. \textbf{b.} The fit residuals for each of the traces track consistently throughout the wavelength sweep. These are cropped to the gray region in the above traces to emphase the smaller perturbations}
    \label{fig:traces}
\end{figure}

Traces of the change in angle for the acquired experimental data are shown in \autoref{fig:traces}a. The wavelength change was acquired for the full 1000\,pm, however each plot was limited to 20\,pm for clarity. The characteristic wavelength change, $\Delta\lambda_0$, was measured to be approximately 1.1\,pm, yielding a analysis range of around $20\Delta_\lambda$. Depicted in the figure are four traces, showing the how the angle changes with differing degrees of uncorrelated and correlated noise. Correlated noise was reduced by subtracting the average of the aforementioned dark frames. Additional uncorrelated noise was sampled from a Gaussian distribution with scale $2^8$ and added to the raw frames to create the added noise traces. Of particular note is the initial and asymptote angle relating to the uncorrelated and correlated noise contributions respectively. Additional uncorrelated noise increases the departure angle. Similarly, correlated noise reduces the asymptote angle. \autoref{fig:traces}b shows the residuals between the data points and the fits in \autoref{fig:traces}a. The flattening of the dark subtracted residual traces in \autoref{fig:traces}b for larger wavelength deviations is due to the speckle patterns approaching orthogonality with the initial reference pattern. Similarly, the greater variations for low perturbations is due to increased sensitivity near zero wavelength change.

\begin{figure*}[htbp]
    \centering
    \includegraphics[width=0.9\textwidth]{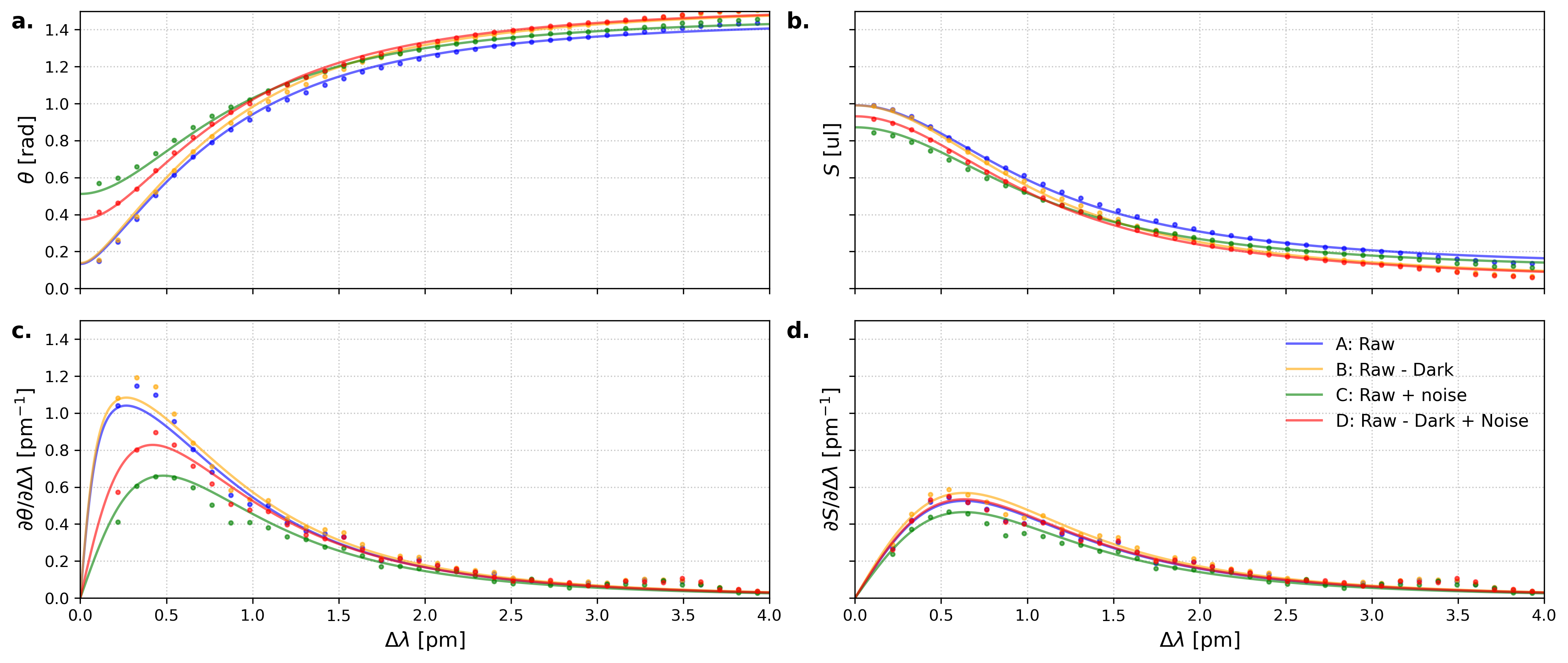}
    \caption{A direct comparison between the angle and similarity metrics are shown. Similar to \autoref{fig:traces}b, these traces are for the small perturbation region where the differences are most pronounced. \textbf{a.}  The angle demonstrates the effect of uncorrelated noise clearly. The raw and dark subtracted frames approach a similar bound, with the noise-added traces separating earlier. \textbf{b.} Compared to a., the low $\Delta\lambda$ values only separate slightly. Crucially, the traces without added noise both appear to approach unity, obfuscating the effects of uncorrelated noise. The non-zero high $\Delta\lambda$ asymptote is similar for both angle and similarity as this is where similarity performs well. \textbf{c.} The derivative of the angle highlights multiple features. The high gradient values biased around zero demonstrate the improved sensitivity, particularly for lower uncorrelated noise contributions. For extremely small $\Delta\lambda$ values, the measurement becomes noise dominated and the gradient drops to zero. \textbf{d.} The gradient of the similarity is largely invariant between traces, with little benefit to reducing uncorrelated noise. Even the lowest noise trace shows less sensitivity to $\Delta\lambda$ than the highest noise angle trace.}
    \label{fig:metrics}
\end{figure*}

The angle and similarity traces can be compared directly to illustrate their differences. \autoref{fig:metrics} shows a comparison between these metrics, including their partial derivatives with respect to the measurand. The angular metric provides a clear illustration of the influence of uncorrelated noise. The raw and dark-subtracted frames converge toward a comparable small limiting value, whereas the traces with added noise diverge at low $\Delta\lambda$, demonstrating a larger noise-dominated regime. In contrast, the traces at low $\Delta\lambda$ exhibit only modest separation. Notably, in the absence of added noise, both traces appear to asymptotically approach unity, thereby masking the impact of uncorrelated noise. The non-zero asymptotic behaviour at high $\Delta\lambda$ is comparable for both the angle and similarity metrics, consistent with the regime in which the similarity measure remains effective. The derivative of the angular metric resolves multiple informative features. Large gradient values centred near, but not at, zero indicate enhanced sensitivity, particularly for smaller contributions of uncorrelated noise. For sufficiently small $\Delta\lambda$, the measurement becomes dominated by uncorrelated noise sources and the gradient correspondingly approaches zero. The gradient of the similarity metric, however, is largely invariant across traces, incorrectly suggesting limited benefit from reducing uncorrelated noise. Even under the lowest-noise condition, the similarity gradient exhibits reduced sensitivity to $\Delta\lambda$ compared with the angular metric under the highest-noise condition. This strongly emphases the efficacy of the angle metric of similarity for small measurand perturbations.

\section{Resolution bound}
Using the speckle angle, it is also possible to estimate the smallest resolvable change in measurand for a given captured speckle pattern. Consider an experiment in which the intensity patterns are captured with bit depth, $b$, and finite pixel count, $N$. In the absence of noise, the smallest possible change in distance corresponds to a single pixel changing by one bit orthogonally. Let us assume that the camera is not saturated and the intensity pattern has mean $\langle I \rangle$ and standard deviation $\sigma_I$. This one bit change can then be normalised as per \autoref{eq:vec_transform}. Geometrically, this corresponds to a change in angle 
\begin{equation}
    \theta_\mathrm{min} = \arctan\left(\frac{1}{\sigma_I\sqrt{N}}\right).
\end{equation} Note, this bound specifies the smallest change in angle that is resolvable, which is not \emph{ipso facto} the true sensitivity of the measurement. The scattering optic and experimental structure will define the sensitivity and the noise sources will introduce potentially more limiting bounds, and thus define the overall noise floor. This bound merely provides the theoretical lower informational limit in the absence of noise.

In the experimental example, the speckle intensity pattern was produced by an integrating sphere. The field is unpolarized and follows the gamma distribution given by \autoref{eq:sphere_pdf}~\cite{goodmanStatisticalPropertiesLaser1975,facchin_wavelength_2021}. If the exposure is tuned as to not oversaturate the camera, the sampled intensity pattern is thus distributed as,

\begin{equation}
    I \sim \frac{4I}{\langle I\rangle}e^\frac{-2I}{\langle I\rangle} \quad \sigma_I = \frac{\langle I \rangle}{\sqrt{2}}
    \label{eq:sphere_pdf}
\end{equation}

The smallest possible speckle distance that may be captured on the camera can then be shown to be $\theta_\mathrm{min} \simeq 1/\sqrt{N}\sigma_I = \sqrt{\frac{2}{N}} \langle I \rangle$. But, for an integrating sphere, $\theta \simeq \sqrt{2}\Delta$. Thus, assuming a competent measurement (i.e. not overexposed), we obtain a resolvability bound.

\begin{equation}
     \Delta_\mathrm{min} \simeq \frac{1}{\sqrt{N}\langle I\rangle}
\end{equation}

From the wavelength experiment, $\Delta\lambda_0 = 1.1\mathrm{pm}$, and the camera capturing $640\times 64$ pixels with a bit depth of 14. The exposure is such that the mean intensity value is 16.93\% saturation, or 2773.6 units of $2^{14}=16384$. Given these values, the minimum speckle angle and wavelength change would be $\theta_\mathrm{min} \simeq 2.519\,\si{\micro rad}$ and $\Delta_\mathrm{min}\simeq 1.781\times10^{-6}$ respectively. Scaling by $\Delta\lambda_0$ yields a resolution limit of 1.960\,am for wavelength changes at $1550\,\si{nm}$. Even though this represents an acquisition bound, knowledge of intrinsic apparatus resolvability decoupled from the measurand has clear value. For comparison, \autoref{fig:traces}a shows the uncorrelated noise of $\theta_{0,A} = 0.133$ rad, corresponding to $\Delta\lambda_\mathrm{min} \simeq \Delta\lambda_0\theta_{0,A}/\sqrt{2} \simeq 103.4$ fm.

Another interesting feature is the measurand perturbation corresponding to maximum sensitivity, as demonstrated in \autoref{fig:metrics}c. The presence of uncorrelated noise the gradient forces the sensitivity to approach zero as $\Delta\rightarrow0$. For the raw dataset this maxima is 1.041 mrad/fm and occurs at $\Delta\lambda=267.1$ fm, or $0.243\Delta\lambda_0$. In contrast, the noise added trace had a maxima of 0.6617 mrad/fm at 484.8 fm, or $0.441\Delta\lambda_0$. This demonstrates that not only does the uncorrelated noise harm the resolution of the measurement, it also reduces the sensitivity when above the noise floor. Although this does not mark an absolute sensitivity bound, it provides additional context when describing the noise floor of the measurement.

If the objective of this experiment was to resolve the smallest possible wavelength changes, reducing uncorrelated intensity noise should be prioritized over further increasing pixels or bit depth.

\section{Conclusions}
In this paper we have introduced the concept of speckle angle, a geometric quantity which emerges naturally from the established similarity formalism used increasingly throughout speckle metrology. Speckle angle provides a physically meaningful measure of intensity pattern evolution that remains linear and maximally sensitive in the small-perturbation regime, where it enables more direct noise floor analysis. A case study was conducted experimentally using wavelength measurement within integrating spheres, demonstrating the influence of both uncorrelated and correlated noise. This approach also inherently provides a common metric for comparing disparate speckle metrology experiments by focusing on the speckle vectors directly. More broadly, the geometric perspective suggests that speckle measurements may be interpreted as trajectories through a high-dimensional space, opening new opportunities for measurand discrimination and multidimensional sensing.

\begin{backmatter}
\bmsection{Funding} We thank the Australian Research Council for funding (grant FL210100099). This research was supported by an Australian Government Research Training Program (RTP) Scholarship. This work is supported by the U.S. Office of Naval Research Global (N62909-25-1-2029)

\bmsection{Acknowledgement} We thank the Australian Research Council for funding (grant FL210100099). This research was supported by an Australian Government Research Training Program (RTP) Scholarship. This work is supported by the U.S. Office of Naval Research Global (N62909-25-1-2029)

\bmsection{Disclosures} The authors declare no conflicts of interest.

\bmsection{Data availability} Data is available from the author upon request.

\bmsection{Supplemental document}
A supplemental document must be called out in the back matter so that a link can be included. For example, “See Supplement 1 for supporting content.” Note that the Supplemental Document must also have a callout in the body of the paper.
\end{backmatter}

\bibliography{refs}

\bibliographyfullrefs{refs}



\end{document}